# Record Maximum Oscillation Frequency in C-face Epitaxial Graphene Transistors

Zelei Guo[1], Rui Dong[1], Partha Sarathi Chakraborty[2], Nelson Lourenco[2], James Palmer[1], Yike Hu[1], Ming Ruan[1], John Hankinson[1], Jan Kunc[1], John D. Cressler[2], Claire Berger[1,3], and Walt A. de Heer[1*]

1. School of Physics, Georgia Institute of Technology, Atlanta GA, 30332, USA
2. School of Electrical and Computer Engineering, Georgia Institute of Technology, Atlanta GA, 30332, USA
3. CNRS-Institut Néel, BP166, 38042 Grenoble Cedex 9, France
*: Corresponding author

**Abstract**
The maximum oscillation frequency ($f_{max}$) quantifies the practical upper bound for useful circuit operation. We report here an $f_{max}$ of 70 GHz in transistors using epitaxial graphene grown on the C-face of SiC. This is a significant improvement over Si-face epitaxial graphene used in the prior high frequency transistor studies, exemplifying the superior electronics potential of C-face epitaxial graphene. Careful transistor design using a high κ dielectric T-gate and self-aligned contacts, further contributed to the record-breaking $f_{max}$.

Graphene is considered a promising candidate material for high-frequency electronics[1]. Whereas the lack of a bandgap in graphene, and consequently the small switching on-off current ratio, clearly impedes the development of graphene-based digital electronics, large on-off ratio is not necessary for some high-frequency circuits. The intrinsic low dimensionality of graphene, high carrier mobility, and large carrier velocity are clear advantages for realizing high frequency performance [2,3]. From the first large scale patterning of graphene field effect transistors (GFET) [4], performance has improved at a rapid pace [2,3]. For RF transistors, the two most important small-signal figures-of-merit are the cut-off frequency $f_T$ (the frequency at which the current gain is unity), and the maximum oscillation frequency $f_{max}$ (the frequency at which the power gain is unity). De-embedded cut-off frequencies in graphene FETs are starting to approach the values of the best of Si-based transistors [5]. Despite the high $f_T$ of graphene FETs, however, transistor power gain ($f_{max}$) remains stubbornly low, limiting its capabilities for use in practical circuits [6]. In any large-scale practical electronics application, graphene is implemented on a substrate that may greatly affect the underlying structural and electronic properties. An atomically well-defined substrate is clearly preferable over disordered substrates for most electronic applications. This has been realized by decades of transistor development on mono-crystalline silicon, where devices that are a few tens of nanometers in size are now reliably patterned by the billions/cm$^2$. Epitaxial graphene on mono-crystalline 4H- or 6H-SiC wafers has a demonstrated large-scale patterning capability [4, 6-9]. Large wafers with atomically smooth surfaces that are ready for epitaxial growth are commercially available (at a price of about US $20/cm$^2$). Because SiC is a wide bandgap semiconductor, epitaxial graphene does not need to be transferred to another substrate. Another advantage is that the graphene-substrate interface is formed at high temperatures, so that it is well defined, pure and reproducible.

On SiC substrates, previous efforts towards high-frequency GFET production have exclusively focused on the (0001) Si-terminated surface [8-10], although the (000-1) carbon terminated surface has reportedly higher electronic mobility [11]. However, monolayer epitaxial graphene is more difficult to produce on the C-face than on the Si-face [12]. Very thin films are required in this case because of the strong potential screening of the top gate by the graphene layers [4, 13].

In a basic GFET structure, the average carrier density (n) in the graphene channel is modulated by the voltage applied on the gate ($V_{gs}$), which is reflected by the induced change in the current $I_{ds}$ that flows in the channel from drain to source, and by the transconductance ($g_m = dI_{ds}/dV_{gs}$) (see for instance [14]). Specifically, $I_{ds}=(W/L).nev$, where L and W are the channel length and width. The average carrier velocity v can be written $v = 1/(1/v_{sat} + 1/\mu E)$, in which $v_{sat}$ is the carrier saturation velocity, $\mu$ is the carrier mobility, and E is the average electric field along the channel. For high frequencies, the current gain is defined by the ratio $|H_{21}|=\partial I_{ds}/\partial I_{gs}$, where $I_{gs}$ is the gate-source current. The cutoff frequency $f_T$ is the frequency at which $|H_{21}|=1$, and the maximum oscillation frequency $f_{max}$ can be derived from measurements as the frequency at which the Mason's unilateral gain $U^{1/2}=1$.

For enhanced performance at high frequencies, higher carrier velocities have been pursued, which can be achieved via higher $v_{sat}$ or $\mu$. For graphene there is no clear evidence for a saturation velocity per se. However, following ref. [14] where the observed $v_{sat}$ is modeled by hot carriers scattering by optical phonons from the substrate ($v_{sat} = v_F(\hbar\Omega/E_F)$, with $\hbar\Omega$ the relevant optical phonon energy), we expect a higher upper bond for $v_{sat}$ for epitaxial graphene on SiC ($\hbar\Omega \sim 115$ meV [15]) than for SiO$_2$ ($\hbar\Omega \sim 55$-60 meV [2, 14]) similarly to graphene transferred to diamond-like carbon ($\hbar\Omega \sim 165$ meV) [16]. We note that $v_{sat}$ could also be affected by the properties of the top-gate dielectric layer. Carrier mobilities up to $\mu \approx 2000$ cm$^2$/Vs have been reported for Si-face epitaxial graphene after an hydrogenation process that results in a significant enhancement in transport properties [17]. For top gated monolayer graphene on the C-face mobility $\mu \approx 8000$ cm$^2$/Vs (Fig. 1b) is routinely achieved at high charge density, and up to 40,000 cm$^2$/Vs at low charge density [11]. High carrier mobility can result in lower contact resistances, which are beneficial for $f_{max}$ [18], and would make it easier to drive the device into saturation. For example, with a mobility $\mu = 2000$ cm$^2$/Vs [17], an electric field 120 kV/cm would be required to accelerate the carriers to 75% of the saturation velocity (if limited by substrate phonons). The electric field can be reduced well below thermal limits by using high mobility C-face graphene with $\mu = 8000$ cm$^2$/Vs. These considerations make high quality monolayer graphene on the C-face of SiC a very good candidate for realizing high frequency FETs.

In this letter, we show that high mobility C-face GFETs built using T-gates and self-aligned contacts indeed can present promising high-frequency characteristics. In particular, their maximum oscillation frequency $f_{max}$ of 70 GHz outperforms published GFETs fabricated on any other substrate materials. $f_T$ is often referred to as the most relevant figure-of-merit for graphene RF transistors; however, $f_{max}$ is a far more relevant parameter for most circuits, since it accounts for device parasitics and measures the frequency at which RF power can be delivered to a load.

Here, high $f_{max}$ have been achieved by a combination of high mobility graphene and FET design. Owing to the high carrier mobility of C-face graphene, the contact resistance between graphene and metal is low (value $R_c$ <100 Ωμm). The un-gated graphene resistance is minimized by utilizing a self-alignment technique, thereby minimizing the total access resistance (sum of contact resistance and un-gated graphene resistance). In addition, the self-alignment technique minimizes parasitic capacitance. T-gate structures are widely used to improve the power gain performance in III-V and silicon RF transistors, and can reduce the gate resistance, especially [19] for short gate lengths, by incorporating a large, conductive gate head.

Monolayer C-face epitaxial graphene is grown on semi-insulating 4H-SiC single crystals using the confinement controlled sublimation method (CCS) [12]. For this approach, a 3 mm by 4 mm SiC die is inductively heated up to 1500 °C in a graphite enclosure provided with a calibrated leak. An example of a monolayer area of several hundreds of μm$^2$ growing over the SiC surface steps is shown in Figure 1. Monolayer graphene regions are identified by their Raman spectroscopy signal combined with atomic force microscopy (AFM) [11, 20]. As shown in Figure1a, the characteristic Raman spectrum of a monolayer graphene region (after subtraction of the SiC Raman peaks) comprises of a narrow G-peak (~1580 cm$^{-1}$) and a single Lorentzian 2D-peak (~2685 cm$^{-1}$). The absence of D-peak indicates that the graphene is of high structural quality. The characteristic half-integer quantum Hall effect is observed in similarly prepared monolayer graphene [11, 12, 20], demonstrating the quality of the graphene monolayer on the C-face, even after deposition of a top-gate [11]. The deposition of a top gate dielectric layer (ALD Al$_2$O$_3$, see below) moderately degrades the graphene mobility. High Hall and FET mobilities are consistently observed on the p- and n- side of ambipolar C- face monolayer graphene devices, as shown for instance in Figure 1b, with $\mu_{Hall}$ = 7500 cm$^2$/Vs and $\mu_{FET}$ = 8700 cm$^2$/Vs on the n-doped side of a 2 μm x 4 μm Hall bar. The mobility $\mu_{FET}$ was calculated using $\mu_{FET}$ = [∂ (Conductivity)/∂V$_{gs}$]/ C$_{ox}$, where C$_{ox}$ is determined from measuring the charge induced by the gate in Hall effect measurements. These values are the highest $\mu_{FET}$ reported so far for epitaxial graphene. Note that this graphene layer is n-doped (n ~ 1.6 x 10$^{12}$/cm$^2$) as expected from charge transfer from the SiC substrate [1]. However, device processing and gate deposition often compensate for the n-doping and may turn the device p-doped (as is the case for the GFET presented below in Figure 4). A schematic diagram of the GFET fabrication process flow is shown in Figure 2.

Standard lithography and metallization techniques were used, and are compatible with large-scale mass production. T-gates of various lengths were first patterned on the selected monolayer graphene regions using e-beam lithography (JEOL JBX 9300), with a tri-layer electron beam resist (Figure 2a) [21]. A thin layer of aluminum (4~5 nm) was deposited as the seed layer for subsequent low temperature atomic layer deposition (ALD) of 10 nm thick aluminum oxide as the dielectric layer (Figure 2b) [22, 23]. E-beam evaporation was used to deposit 40 nm Ti / 130 nm Au on the dielectric to complete the gate stack (Figure 2c), and the remaining metal was lifted off (Figure 2d). Unlike for most wafer-scale graphene FET fabrication processes reported in the literature [4, 8, 9] that involve covering the entire wafer with a dielectric layer, the above method applies the dielectric layer only to the gated region of the devices. The remaining area of graphene is left uncovered for the subsequent self-aligned contact application. The source and drain areas were coated with 7 nm Pd/10 nm Au by angle deposition, thereby minimizing the exposed graphene area on both sides of the gate foot (Figure 2e). Residual graphene is removed with oxygen-plasma before e-beam lithographically defined source/drain/gate pads are covered with metal (40 nm Ti / 140 nm Au) (Figure 2f). For improved performance, the GFETs have a 2-finger design, with channel width 7 μm, and probing pads of 40 μm x 50 μm with a 100 μm pitch [24]. Figure 3a shows an example of a T-gate with 100 nm foot length made on epitaxial graphene, and Figure 3b is a scanning electron microscopy image of the dual gate device.

The DC performance of the GFET devices with 100 nm and 250 nm gate lengths, respectively, are shown in Figure 4. From the output characteristics of the 100 nm gate device at low drain-source bias, we find that the unit width two-point resistance is only $R_{2pt}$ = 200 Ωμm, indicating that the contact resistance ($R_c = 1/2(R_{2pt} - R_{channel})$) is less than 100 Ωμm. This is the smallest per unit width 2-point resistance for graphene FETs reported to date. The contact resistance is smaller than $R_c$ =230 Ωμm (110 Ωμm, resp.) reported[18] for metal–graphene junction resistance at 300K (6K, resp). It is of similar value to the ultra-low resistance ohmic contacts ($R_c$ below 100 Ωμm) reported in Ref [25] and it is comparable to that of silicon [26]. Because of this small contact resistance, a large current modulation ($I_{ds}$ vs. $V_{gs}$) is observed. For instance, a current density as large as 2.6 mA/μm is observed at $V_{ds}$=-0.8 V and $V_{gs}$=0 V. As is usual for graphene devices, the I-V characteristics of the devices doesn't show a clear saturation, but rather an inflection point at relatively small $V_{ds}$ [27-30]. The I-V characteristics also indicate that the device is heavily p-doped, since a minimum in the $I_{ds}$ vs $V_{ds}$ is not reached in the 0-3 V range. In this limited $V_{ds}$ range, large current values were measured, with $I_{ds}$ >2.5 mA/μm at $V_{ds}$=-0.8 V ($g_m$ peaks ~ 250 mS/mm), in the same range as reported in the literature for comparable bias voltages [27-29].

High-frequency scattering parameters (S-parameters) of the GFETs have been measured up to 50 GHz under ambient conditions using standard ground-signal-ground (GSG) microwave probes. The system was calibrated with the short-open-load-through (SOLT) process to eliminate the parasitic effects of the wiring and the probes [8]. Before de-embedding, a maximum $f_T$ of 41 GHz is measured for a device with 100 nm gate length at $V_g$=3.5 V and $V_{ds}$=-0.5 V (Figure 5a). This $f_T$ value rivals the highest non de-embedded cut-off frequency reported to date for GFETs (e.g., comparable to[24]). Measurements on devices of 250 nm gate length show a cut-off frequency of 33 GHz (non de-embedded), demonstrating the reproducibility of high performance of C-face graphene GFETs. To remove the effects of probing pads, we use a de-embedding procedure similar to that described in[16]. Namely, the 'open' and 'short' structures have the exactly same layout as the graphene device under test (DUT), except that there is no graphene channel in the 'open' structure, and all the contacts are shorted to each other in the 'short' structure. After the de-embedding process, $f_T$ values of 110 GHz and 60 GHz were measured on 100 nm and 250 nm gate length devices, respectively. These measured $f_T$ values compare well with the estimated $f_T = g_m/2\pi C_g$ ~ 80-110 GHz using the measured $g_m$ =0.25mS/μm for the 100nm long and 7μm wide dual gate device, with a 15nm thick dielectric and dielectric constant κ = 6~8 for ALD-$Al_2O_3$. As expected, longer gate length devices have reduced $f_T$. Note that considering devices of similar gate length, these $f_T$ values are higher than that of silicon MOSFETs, and about half that of HEMTs.

As discussed, a more relevant parameter for high frequency operation is $f_{max}$. We have measured $f_{max}$ = 38 GHz and $f_{max}$ = 70 GHz before and after de-embedding for a 100 nm gate device. To our knowledge, these $f_{max}$ values are the highest reported values to date for any GFETs, on any substrate [2, 6, 30]. High $f_{max}$ values are consistently found in our C-face graphene FETs. All six of the measured 100 nm gate length devices show $f_{max}$ after de-embedding between 50 GHz and 70 GHz. Data for two of the measured 100 nm devices are shown in Figure 5c. Together, these results show that SiC is a promising substrate for high frequency GFETs.

In prior reports on graphene RF transistors, $f_{max}$ was not only relatively small compared with $f_T$ of the same devices, it was even smaller than that of silicon CMOS with similar gate lengths. The small $f_{max}$ was attributed to large gate resistance, large graphene/metal contact resistances, and non-optimized designs. By contrast, the RF transistors presented here and based on C-face graphene show comparable (high) $f_T$ and $f_{max}$ values. A balanced $f_T$ and $f_{max}$ represents an optimal situation for many RF circuits. These reproducible large $f_{max}$ values result from the T-gate design, the self-alignment technique, and the inherent small contact resistance in the devices.

Due to the T-gate geometry, the gate resistance in this work is on the order of 1 $\Omega/\mu m$ of gate width, an order of magnitude smaller than that without a T-gate, and over two orders of magnitude smaller than the highly doped nanowire gates [24, 28, 31].

The superior performance of our GFETs is also in part due to our self-alignment process that is different from that shown in previous reports [28, 31, 32]. The source and drain contacts are aligned close to the gate foot, instead of being aligned to the gate head. The alignment accuracy (10 nm) is determined by the accuracy in length and the edge roughness of the gate head. This self-alignment method minimizes the ungated portion of the graphene channel, thereby reducing the access resistance and increasing the maximum current density. Parasitic effects that could arise from the contact metal under the overhanging gate head [28] are minimized with a high gate foot (70nm). The parasitic capacitance induced by the T-gate is therefore no more than 10% the gate capacitance of the device made with high-κ dielectric (excluding probe pads and connections), even for 50nm gate lengths.

Contact resistances $R_c$ are also limiting factor for high frequency performance. In previously reported high frequency GFETs, the $R_c$ ranges from ~ 300 $\Omega\mu m$ to several k$\Omega\mu m$ [2]. While the effect of $R_c$ on $f_T$ has been extensively discussed in the context of GFET, its effects on $f_{max}$ is often ignored, even though large $R_c$ degrades $f_{max}$ more than $f_T$. This is because with large $R_c$ and small channel resistance most of the input power is dissipated by the contacts instead of the channel. The absence of current saturation in graphene exacerbates this effect [32], since in the case of GFET there is no mechanism for the channel resistance to become much larger than the contact resistance. In contrast for $f_T$, by driving the GFET at higher $V_{ds}$, a peak transconductance can be achieved, similar to that in small $R_c$ devices.

The small contact resistance $R_c$ between the Pd/Au layer and C-face epitaxial graphene is the key to the record high $f_{max}$ (both with and without de-embedding) reported here. Since $R_c$ < 100 $\Omega\mu m$, at its best operation point, most of the RF input power can be amplified in the graphene channel itself. Following ref [3], $f_{max}$ can be calculated from

$$f_{max} = \frac{f_T}{2\sqrt{g_D(R_G + R_{SD})} + 2\pi f_T R_G C_G},$$ where the channel conductance $g_D \approx 30$mS, source-drain access resistance $R_{SD} \approx 15\Omega$ (due to the self-alignment, the source-drain access resistance comprises of primarily contact resistance on the source and drain), gate resistance $R_G \approx 3\Omega$ and gate capacitance $C_G \approx 5$fF are estimated for a dual 7μm wide T-gate, with the values discussed above. According to the above equation, we find $f_{max}$ =74 GHz for the 100 nm gate length device, in remarkable agreement with the measured $f_{max}$. Because of the low gate resistance, $f_{max}$ is primarily determined by $f_{max} \approx \frac{f_T}{2\sqrt{g_D(R_G + R_{SD})}}$. A low contact resistance together with a low gate resistance is thus key to the comparable and high values of $f_T$ and $f_{max}$.

In summary, we have demonstrated a new fabrication method for RF transistors using high mobility C-face epitaxial graphene for the first time. With T-gate and self-alignment techniques to minimize gate and access resistance, record high $f_{max}$ of 38 GHz and 70 GHz, before and after de-embedding, respectively, are observed on GFETs with 100 nm gate length. Reproducible results on multiple devices show comparably high $f_T$ and $f_{max}$, which is ideal for implementation in high frequency circuits. Our studies open a pathway to better power gain performance in GFETs, and show a great potential of C-face epitaxial graphene in high frequency applications.

**Acknowledgement**
This research was supported by the W. M. Keck Foundation, the AFSOR grant No. FA9550-10-1-0367, and the NSF MRSEC Program under Grant No. DMR-0820382. Zelei Guo would like to thank Thomas J. Beck for helpful discussions.

# Figure captions

Figure 1
(a) Raman spectra of C-face monolayer graphene before (black) and after (red) subtraction of SiC Raman peaks, showing the G and 2D graphene peaks. Note that the D peak is very small. (b) Conductivity versus top gate voltage on C-face monolayer graphene 2 μm x 4 μm Hall bar. On the n-doped side, Hall mobility $\mu_{Hall}$ = 7500 cm$^2$/Vs at n = 1.6x10$^{12}$ / cm$^2$ ($V_g$ = 0V), and FET mobility is $\mu_{FET}$ = 8700 cm$^2$/Vs. On the p-side, $\mu_{FET}$ ~5000 cm$^2$/Vs. Inset: AFM image of a C-face monolayer graphene (scale bar 5μm) over SiC steps. The white lines are graphene pleats characteristic of C-face graphene.

Figure 2
Process flow for GFET fabrication. (a) Monolayer graphene on C-face of SiC. (b) The T-gate is patterned using tri-layer resist and e-beam lithography, followed by aluminum seed layer deposition and ALD of Al$_2$O$_3$. (c) Ti/Au is deposited on top as the gate metal. (d) After lift-off, the T-gate stands on graphene. Note the sides of the T-gate coated by an insulating layer. (e) Angle deposition of Pd/Au to form self-aligned contacts. (f) Ti/Au source and drain are deposited on top of the self-aligned contacts as probing pads.

Figure 3
(a) Scanning electron microscope (SEM) image of a T-gate with 100nm gate foot fabricated with a tri-layer resist, and source and drain contact metal aligned to the gate foot (scale bar 100nm). (b) SEM image of a dual-gate GFET on C-face SiC (scale bar 1μm). Source (S), drain (D) and gate (G) are indicated.

Figure 4
DC characteristics of GFETs on C-face graphene with 100 nm and 250 nm gate lengths. (a)-(b) Current density plotted as a function of gate voltage at $V_{ds}$ =-0.1V and -0.5V for gate length (a) 100nm and (b) 250nm. (c)-(d) Drain-source IV characteristics at gate voltage ranging from 0V to 3V at 0.5V step on (c) 100nm and (d) 250nm gate devices. Maximum current density of ~ 2.6mA/μm is observed.

Figure 5
High frequency characteristics of GFETs on C-face graphene with different gate lengths. (a) $H_{21}$ versus frequency on two devices with 100 nm gate length; device 1 (red) before de-embedding (Solid red line) and after de-embedding (red circles); device 2 (blue, ibidem). Cutoff frequency (by extrapolation of theoretical slope of 20dB/decade to $|H_{21}|$=1) for device 1 and 2 are before de-embedding $f_T$= 41 GHz (31GHz), resp, and after de-embedding $f_T$=110 GHz (90 GHz), resp. (b) $|H_{21}|$ for a 250nm gate length GFET before (solid line; $f_T$= 32 GHz) and after (circles, $f_T$= 60 GHz) de-embedding. (c) Mason's unilateral gain versus frequency for the two 100 nm gate GFETs shown in (a). Intercepts $U^{1/2}$=1 (slope 20dB/dec) give $f_{max}$ =38 GHz and 33 GHz before and 70 GHz after de-embedding. (d) Same as (c) for the 250nm gate length GFET in (b) $f_{max}$=36 GHz and 58 GHz before and after de-embedding.

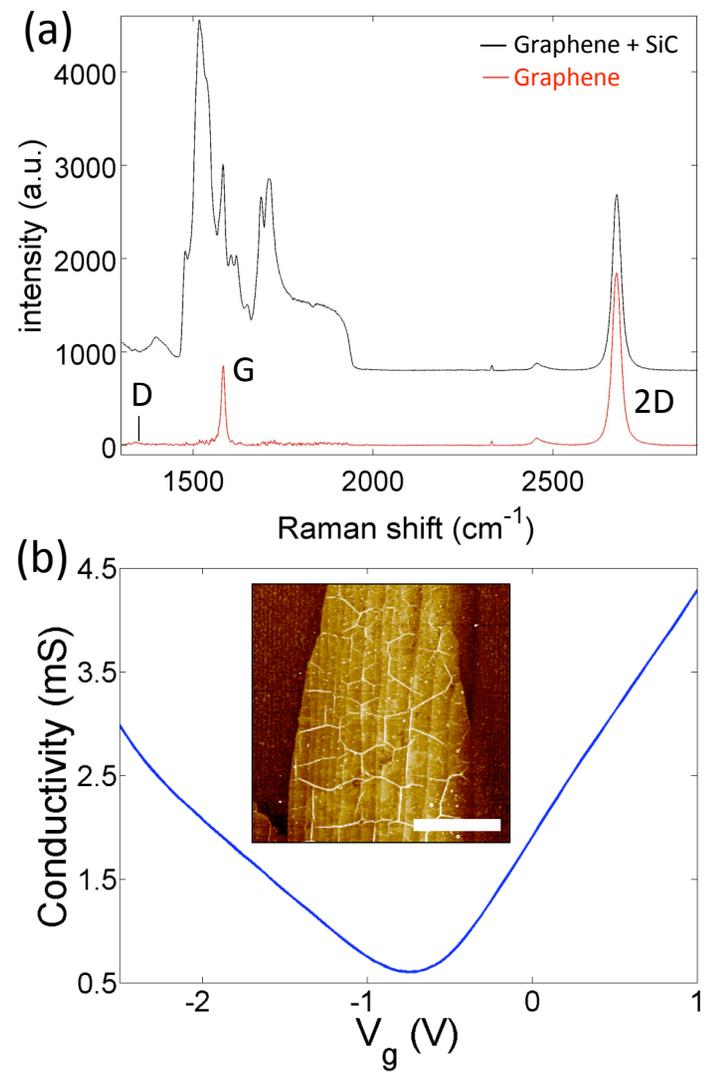

Fig. 1

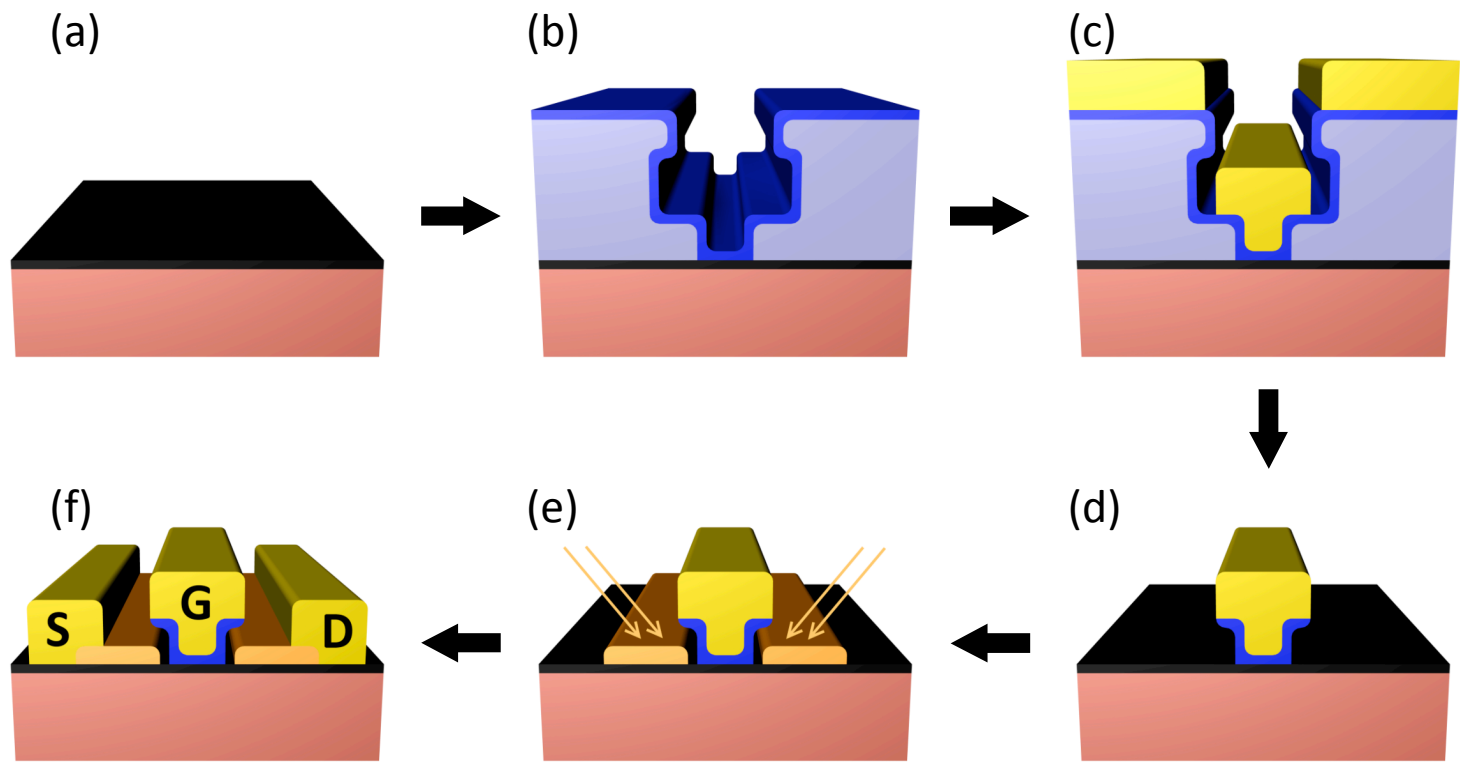

Fig. 2

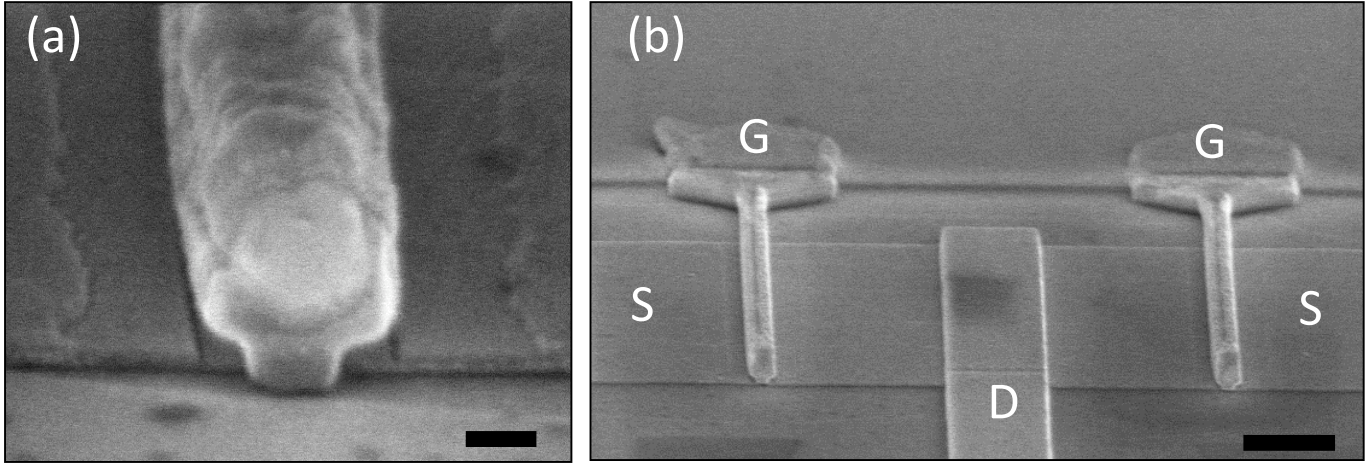

Fig. 3

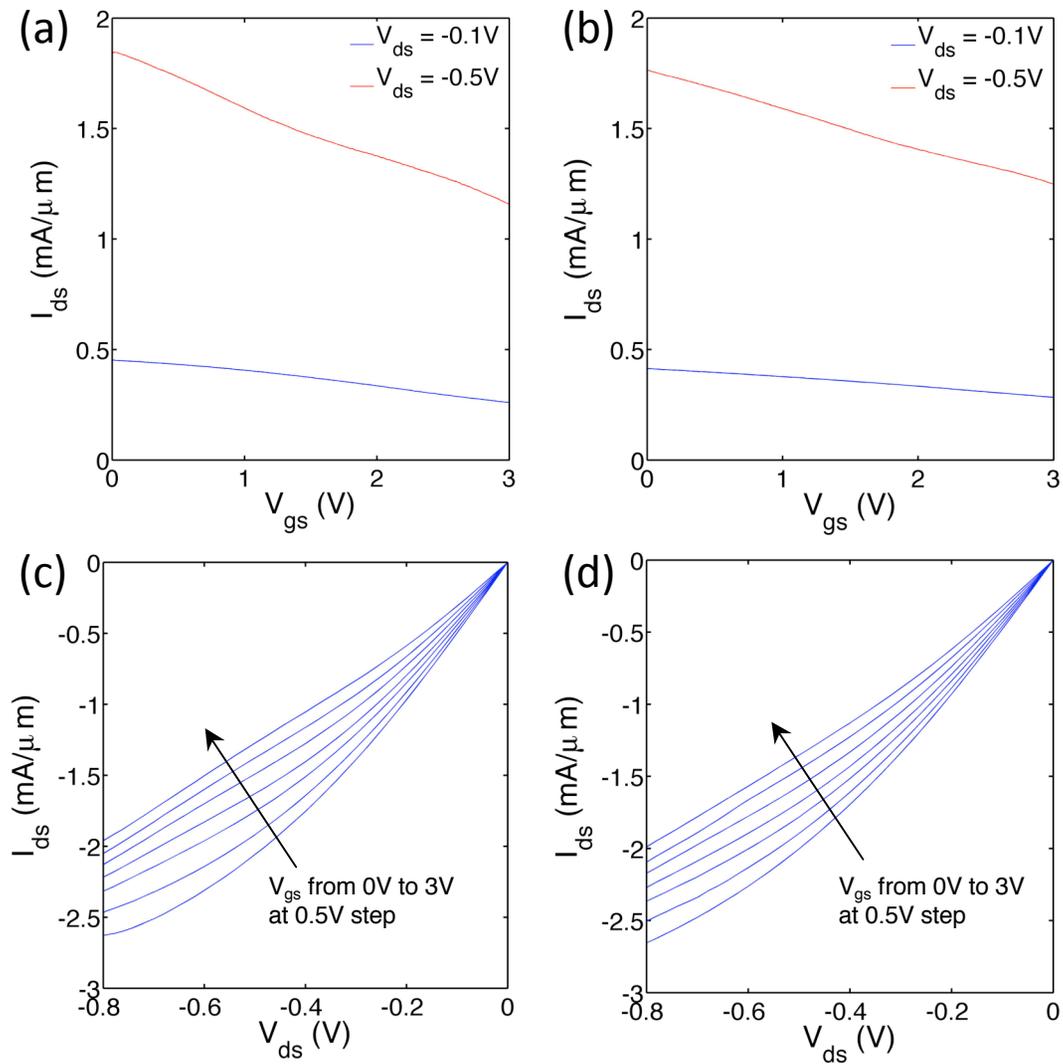

Fig. 4

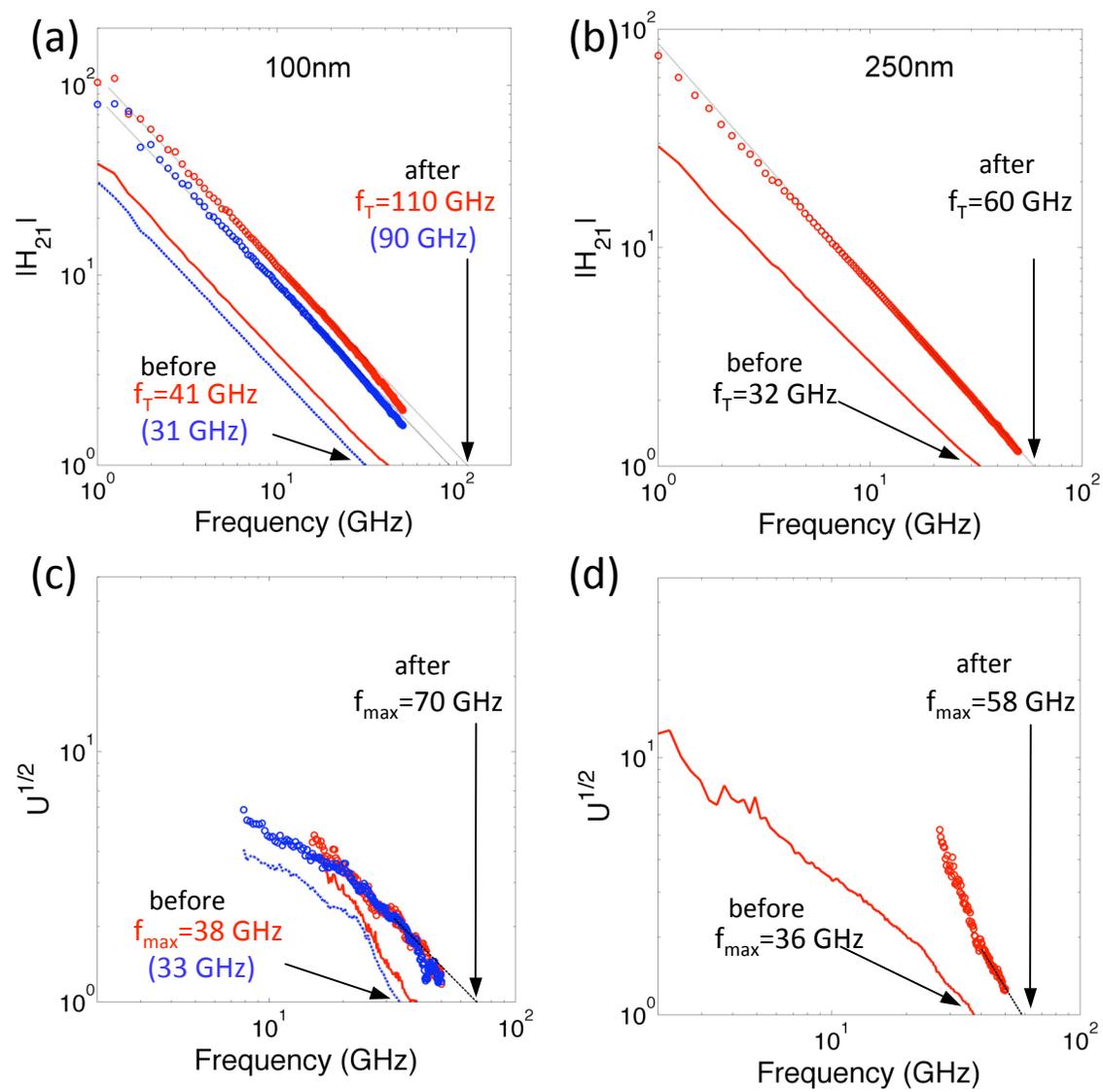

Fig. 5